\documentclass[journal]{IEEEtran}
\PassOptionsToPackage{table}{xcolor}

\usepackage{lettrine}
\usepackage[subtle]{savetrees}
\usepackage{hyphenat}
\usepackage{url}
\usepackage{latexsym}
\usepackage{amsmath}
\usepackage{amssymb}

\usepackage{amsthm}
\usepackage{graphicx}
\usepackage{booktabs}
\usepackage{tabularx}
\usepackage{placeins}
\usepackage[ruled,vlined,linesnumbered]{algorithm2e}
\usepackage{tikz}
\usetikzlibrary{arrows.meta,positioning,fit,backgrounds}
\graphicspath{{figs/}}

\usepackage{xcolor}
\definecolor{linkblue}{RGB}{0,51,204}
\usepackage[colorlinks=true,
linkcolor=linkblue,
citecolor=linkblue,
urlcolor=linkblue,
filecolor=linkblue]{hyperref}

\definecolor{f4pblue}{HTML}{0F4D92}
\definecolor{f4pblue2}{HTML}{3775BA}
\definecolor{f4pteal}{HTML}{42949E}
\definecolor{f4pred}{HTML}{B64342}
\definecolor{f4pviolet}{HTML}{9A4D8E}
\definecolor{tabhead}{HTML}{DCE7F5}
\definecolor{tabzebra}{HTML}{F3F7FC}

\newcommand{\tablestyle}{%
  \ifdefined\markeverypar\markeverypar{\the\everypar}\fi
  \setlength{\aboverulesep}{0pt}%
  \setlength{\belowrulesep}{0pt}%
  \setlength{\heavyrulewidth}{0.9pt}%
  \setlength{\lightrulewidth}{0.4pt}%
  \arrayrulecolor{f4pblue}%
  \renewcommand{\arraystretch}{1.3}%
  \rowcolors{2}{white}{tabzebra}}
\newcommand{\tabheadrow}{\rowcolor{tabhead}}
\newcommand{\tabh}[1]{\textcolor{f4pblue}{\bfseries #1}}

\tikzset{
  stage/.style={rectangle, rounded corners=3pt, draw=f4pblue, line width=0.6pt,
                fill=f4pblue!6, align=center, inner sep=3pt,
                text width=0.80\columnwidth, font=\scriptsize},
  badge/.style={circle, fill=f4pblue, text=white, inner sep=0pt,
                minimum size=3.8mm, font=\scriptsize\bfseries},
  gram/.style={rectangle, rounded corners=2.5pt, line width=0.6pt, align=center,
               inner sep=2pt, text width=0.15\columnwidth, font=\scriptsize},
  flow/.style={-{Stealth[length=3.6pt,width=3pt]}, draw=f4pblue!85, line width=0.6pt},
  loopbox/.style={draw=f4pblue!55, dashed, line width=0.5pt, rounded corners=5pt,
                  fill=f4pblue!2},
}

\makeatletter
\let\hp@oldeqnnum\@eqnnum
\def\@eqnnum{{\color{linkblue}\hp@oldeqnnum}}
\makeatother

\theoremstyle{definition}
\newtheorem{problem}{Problem}
\newtheorem{proposition}{Proposition}

\makeatletter
\renewcommand{\SetKwInput}[2]{%
	\algocf@newcommand{#1}[1]{%
		\sbox\algocf@inputbox{\hbox{\KwSty{#2} }}%
		\ifthenelse{\boolean{algocf@hanginginout}}{\relax}{\algocf@seteveryparhanging{\relax}}%
		\ifthenelse{\boolean{algocf@inoutnumbered}}{\relax}{\algocf@seteveryparnl{\relax}}%
		{\let\\\algocf@newinput\hangindent=\wd\algocf@inputbox\hangafter=1\unhbox\algocf@inputbox##1\par}%
		\algocf@linesnumbered%
		\ifthenelse{\boolean{algocf@hanginginout}}{\relax}{\algocf@reseteveryparhanging}%
}}%
\makeatother
\SetKwInput{KwIn}{Input}
\SetKwInput{KwOut}{Output}
\SetAlgoCaptionSeparator{.}
\SetAlFnt{\small}
\SetAlCapFnt{\small}
\SetAlCapNameFnt{\small}

\begin{document}

\title{\textit{Ranking the Parameters Shaping Grid Dynamics}}

\author{Ahmad~F.~Taha, \textit{Member, IEEE}%
\thanks{A. F. Taha is with Civil and Environmental Engineering, Vanderbilt
University, Nashville, TN 37203, and is also affiliated with the Department of
Electrical and Computer Engineering at Vanderbilt University (e-mail
\url{ahmad.taha@vanderbilt.edu}). This work is supported by the National
Science Foundation (NSF), United States, under grants 2152450 and 2550934.}}

\maketitle

\begin{abstract}
Generator and transmission line parameters do not shape the time-domain behavior of a power system equally; a parameter that dominates one dynamic property can be almost invisible to another. This paper studies how the physical parameters of a detailed and general nonlinear differential algebraic equation (DAE) grid model shape the properties that decide whether system states such as frequencies, voltages, and angles can be inferred, observed, controlled, and kept stable. The rankings are conditional as they depend on where grid sensors and actuators sit, on the operating region and the disturbance, on the network topology, and on other settings that the operator chooses. The framework combines DAE trajectory sensitivities, variational Gramians, and stability margin sensitivities to measure how much each parameter reshapes the network dynamics. On three benchmark systems, the rankings show where dynamic leverage concentrates and where the control-theoretic metrics disagree, and they point to how parameter studies can guide calibration, sensor and actuator placement, and security assessment. The proposed methods herein are also scalable and can be tested on grids with tens of thousands of buses. 
\end{abstract}


\section{Introduction and Paper Contributions}

\lettrine{W}{HEN} a fault, a load swing, or a drop in renewable output disturbs a power system, the measured response reflects many physical quantities at once. Generator inertia and damping shape it, and so do machine reactances and time constants, controller settings, load behavior, and the network power flow parameters that enter through the admittance matrix. A few of these quantities dominate the visible response. Others have little effect, are hard to tell apart, respond weakly to the available actuators, or show up in only one mode or one contingency. Knowing which parameters carry the most weight, and for which purpose, helps focus model calibration and guide sensor and actuator placement, and it cuts the computational burden of dynamic security studies. We study this problem as an \emph{offline} assessment of parameter importance. It belongs to planning, calibration, and model validation ahead of operation, so the design goal is thoroughness across many scenarios, and online speed is secondary.

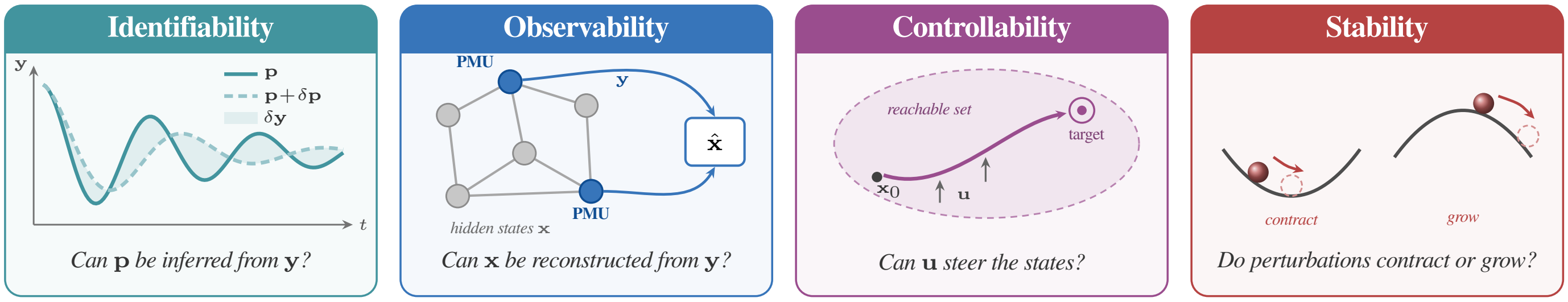
\begin{figure*}[t]
\centering
\resizebox{\textwidth}{!}{%
\begin{tikzpicture}[font=\scriptsize, >={Stealth[length=3.2pt,width=2.8pt]},
  frame/.style={rounded corners=5pt, line width=0.6pt},
  ttl/.style={font=\small\bfseries, text=white, anchor=center},
  ask/.style={font=\scriptsize\itshape, text=black!80, anchor=center},
  note/.style={font=\tiny\itshape}]
\begin{scope}
\fill[f4pteal!5, rounded corners=5pt] (0,0) rectangle (3.9,3.05);
\fill[f4pteal, rounded corners=5pt] (0,2.55) rectangle (3.9,3.05);
\fill[f4pteal] (0,2.55) rectangle (3.9,2.75);
\draw[frame, f4pteal] (0,0) rectangle (3.9,3.05);
\node[ttl] at (1.95,2.8) {Identifiability};
\draw[->, black!55, line width=0.5pt] (0.3,0.75) -- (3.65,0.75) node[right, font=\tiny, text=black!70, inner sep=1.5pt] {\(t\)};
\draw[->, black!55, line width=0.5pt] (0.3,0.75) -- (0.3,2.42) node[left, font=\tiny, text=black!70, inner sep=1.5pt] {\(\mathbf{y}\)};
\fill[f4pteal!16] plot[domain=0:3.15, samples=120] ({\x+0.4},{1.5+0.72*exp(-0.55*\x)*cos(5.5*\x r)})
  -- plot[domain=3.15:0, samples=120] ({\x+0.4},{1.5+0.72*exp(-0.85*\x)*cos(4.2*\x r)}) -- cycle;
\draw[f4pteal, line width=1.1pt, domain=0:3.15, samples=120, shift={(0.4,1.5)}]
  plot (\x,{0.72*exp(-0.55*\x)*cos(5.5*\x r)});
\draw[f4pteal!55, line width=1.1pt, dash pattern=on 2.6pt off 1.6pt, domain=0:3.15, samples=120, shift={(0.4,1.5)}]
  plot (\x,{0.72*exp(-0.85*\x)*cos(4.2*\x r)});
\draw[f4pteal, line width=1.1pt] (2.3,2.33) -- (2.65,2.33) node[right, font=\tiny, text=black!80, inner sep=1.5pt] {\(\mathbf{p}\)};
\draw[f4pteal!55, line width=1.1pt, dash pattern=on 2.6pt off 1.6pt] (2.3,2.1) -- (2.65,2.1) node[right, font=\tiny, text=black!80, inner sep=1.5pt] {\(\mathbf{p}+\delta\mathbf{p}\)};
\fill[f4pteal!16] (2.3,1.81) rectangle (2.65,1.93);
\node[right, font=\tiny, text=black!80, inner sep=1.5pt] at (2.65,1.87) {\(\delta\mathbf{y}\)};
\node[ask] at (1.95,0.36) {Can \(\mathbf{p}\) be inferred from \(\mathbf{y}\)?};
\end{scope}
\begin{scope}[xshift=4.15cm]
\fill[f4pblue2!5, rounded corners=5pt] (0,0) rectangle (3.9,3.05);
\fill[f4pblue2, rounded corners=5pt] (0,2.55) rectangle (3.9,3.05);
\fill[f4pblue2] (0,2.55) rectangle (3.9,2.75);
\draw[frame, f4pblue2] (0,0) rectangle (3.9,3.05);
\node[ttl] at (1.95,2.8) {Observability};
\coordinate (a) at (0.5,1.9); \coordinate (b) at (1.15,2.25); \coordinate (c) at (1.3,1.5);
\coordinate (d) at (0.6,1.05); \coordinate (e) at (1.95,2.0); \coordinate (f) at (2.0,1.1);
\draw[black!35, line width=0.7pt] (a)--(b) (a)--(d) (b)--(c) (c)--(d) (b)--(e) (c)--(f) (e)--(f) (d)--(f);
\foreach \n in {a,c,d,e} \filldraw[fill=black!22, draw=black!45, line width=0.5pt] (\n) circle (0.12);
\foreach \n in {b,f} \filldraw[fill=f4pblue2, draw=f4pblue, line width=0.6pt] (\n) circle (0.12);
\node[font=\tiny\bfseries, text=f4pblue, anchor=south east, inner sep=1pt] at (1.02,2.36) {PMU};
\node[font=\tiny\bfseries, text=f4pblue, below=1.3pt] at (f) {PMU};
\node[draw=f4pblue2, fill=white, rounded corners=2pt, line width=0.7pt, minimum width=0.62cm, minimum height=0.5cm] (est) at (3.3,1.62) {\(\hat{\mathbf{x}}\)};
\draw[->, f4pblue2, line width=0.8pt] (b) ++(0.13,0.02) to[out=5, in=120] node[pos=0.45, below, font=\tiny, text=f4pblue, inner sep=1.2pt] {\(\mathbf{y}\)} (est.north);
\draw[->, f4pblue2, line width=0.8pt] (f) ++(0.13,0) to[out=0, in=-120] (est.south);
\node[note, text=black!60] at (1.05,0.72) {hidden states \(\mathbf{x}\)};
\node[ask] at (1.95,0.36) {Can \(\mathbf{x}\) be reconstructed from \(\mathbf{y}\)?};
\end{scope}
\begin{scope}[xshift=8.3cm]
\fill[f4pviolet!5, rounded corners=5pt] (0,0) rectangle (3.9,3.05);
\fill[f4pviolet, rounded corners=5pt] (0,2.55) rectangle (3.9,3.05);
\fill[f4pviolet] (0,2.55) rectangle (3.9,2.75);
\draw[frame, f4pviolet] (0,0) rectangle (3.9,3.05);
\node[ttl] at (1.95,2.8) {Controllability};
\fill[f4pviolet!14] (2.0,1.6) ellipse (1.6 and 0.78);
\draw[f4pviolet!70, dash pattern=on 2pt off 1.5pt, line width=0.5pt] (2.0,1.6) ellipse (1.6 and 0.78);
\node[note, text=f4pviolet!80!black] at (1.4,1.97) {reachable set};
\draw[->, f4pviolet, line width=1.1pt] (0.92,1.27) .. controls (1.6,1.05) and (2.2,1.8) .. (2.84,1.92);
\draw[->, black!60, line width=0.55pt] (1.51,0.98) -- (1.51,1.22);
\draw[->, black!60, line width=0.55pt] (1.99,1.2) -- (1.99,1.46);
\node[font=\tiny, text=black!75] at (1.76,1.06) {\(\mathbf{u}\)};
\draw[f4pviolet, line width=0.6pt] (3.0,1.95) circle (0.13);
\fill[f4pviolet] (3.0,1.95) circle (0.05);
\node[font=\tiny, text=f4pviolet!80!black] at (3.05,1.68) {target};
\fill[black!75] (0.85,1.25) circle (0.055);
\node[font=\tiny, text=black!80] at (0.98,1.1) {\(\mathbf{x}_0\)};
\node[ask] at (1.95,0.36) {Can \(\mathbf{u}\) steer the states?};
\end{scope}
\begin{scope}[xshift=12.45cm]
\fill[f4pred!5, rounded corners=5pt] (0,0) rectangle (3.9,3.05);
\fill[f4pred, rounded corners=5pt] (0,2.55) rectangle (3.9,3.05);
\fill[f4pred] (0,2.55) rectangle (3.9,2.75);
\draw[frame, f4pred] (0,0) rectangle (3.9,3.05);
\node[ttl] at (1.95,2.8) {Stability};
\draw[black!70, line width=1.0pt, domain=-0.72:0.72, samples=40] plot ({1.05+\x},{1.05+0.95*\x*\x});
\draw[f4pred!60, dash pattern=on 1.5pt off 1pt, line width=0.5pt] (1.05,1.16) circle (0.11);
\shade[ball color=f4pred!85] (0.70,1.30) circle (0.11);
\draw[->, f4pred, line width=0.8pt] (0.86,1.46) .. controls (1.0,1.36) .. (1.18,1.31);
\node[note, text=f4pred] at (1.05,0.8) {contract};
\draw[black!70, line width=1.0pt, domain=-0.72:0.72, samples=40] plot ({2.85+\x},{1.95-0.95*\x*\x});
\shade[ball color=f4pred!85] (3.066,2.023) circle (0.11);
\draw[f4pred!60, dash pattern=on 1.5pt off 1pt, line width=0.5pt] (3.533,1.68) circle (0.11);
\draw[->, f4pred, line width=0.8pt] (3.22,2.1) .. controls (3.45,2.02) .. (3.6,1.86);
\node[note, text=f4pred] at (2.85,0.8) {grow};
\node[ask] at (1.95,0.36) {Do perturbations contract or grow?};
\end{scope}
\end{tikzpicture}}
\caption{\normalfont Conceptual view of the four properties in Tab.~\ref{tab:concepts}.}
\label{fig:properties}
\vspace{-0.4cm}
\end{figure*}

The paper poses the question for a general nonlinear differential algebraic equation (DAE) power system model that couples generator dynamics of arbitrary order with the algebraic network and power balance equations \cite{SauerPaiChow2017}. It focuses on the global impact of parameters on four properties, namely identifiability, observability, stability, and controllability. Tab.~\ref{tab:concepts} defines each one, and Fig.~\ref{fig:properties} sketches the question behind each.

\begin{table}[t]
\centering
\caption{\normalfont Control-theoretic properties that this paper studies.}
\label{tab:concepts}
\scriptsize
\tablestyle
\renewcommand{\arraystretch}{1.22}
\setlength{\tabcolsep}{3pt}
\begin{tabularx}{\columnwidth}{@{}>{\raggedright\arraybackslash}p{0.215\columnwidth}>{\raggedright\arraybackslash}X>{\raggedright\arraybackslash}p{0.33\columnwidth}@{}}
\toprule
\tabheadrow \tabh{Property} & \tabh{Central question} & \tabh{Principal dependencies} \\
\midrule
Identifiability & Can parameters be inferred from measured trajectories? & Excitation, sensors, noise, scaling, horizon, topology \\
Observability & Can internal states be reconstructed? & Sensors, trajectory, topology, measurement quality \\
Controllability & Can selected actuators influence relevant states? & Actuator locations, limits, horizon, trajectory, topology \\
Stability & Do feasible perturbations contract or grow? & Trajectory, disturbance, topology, protection, stability metric \\
\bottomrule
\end{tabularx}
\vspace{-0.54cm}
\end{table}

\vspace{0.1cm}

\noindent\textbf{Nonlinear DAE Theory.}
Nonlinear systems theory gives local tests for observability through Lie derivatives and related rank conditions \cite{HermannKrener1977}. In a DAE the dynamic states evolve together with algebraic variables that must satisfy the network equations. \cite{Terrell1997} gives a local observability test for nonlinear DAEs, and \cite{Sato2014} studies controllability and observability for nonlinear index-one DAEs. More recently, \cite{MontanariEtAl2024} propose a rank-based identifiability test that acts directly on nonlinear DAEs. These results lay the theoretical groundwork. Symbolic tests, however, become hard to carry out on large multimachine models, and they usually return a local yes-or-no answer instead of a practical ranking of physical parameters.

\vspace{0.1cm}

\noindent\textbf{Observability of Power System Dynamics.}
The study \cite{DafisNwankpa2002} formulate nonlinear observability directly for a power system DAE with generator dynamics. The work \cite{RouhaniAbur2017} apply Lie derivative tests to a nonlinear fourth-order synchronous generator and find that observability depends strongly on the chosen measurements and on the operating trajectory. For larger systems, the authors in \cite{QiSunKang2015} place PMUs with empirical observability Gramians and our recent work~\cite{kazma2024revisiting} revisits this problem by generalizing it and considering DAE dynamics. An empirical Gramian perturbs the system, simulates it again, and records how strongly states, inputs, or parameters move the trajectories and outputs of the nonlinear system.

\vspace{0.1cm}

\noindent\textbf{Controllability and Stability.}
Control analysis of power systems is mature for models linearized around an operating point. Selective modal analysis, residues, participation factors, and eigenvalue sensitivities are long-standing tools for finding the critical machines, feedback signals, actuator locations, and parameters behind poorly damped modes \cite{VergheseEtAl1982,WangEtAl2000}. Gramian methods and empirical controllability and observability covariances carry this input-output view over to large nonlinear power system models \cite{QiEtAl2017}, and Gramian-based reduction applies directly to large sparse descriptor (DAE) power system models \cite{FreitasEtAl2008}. Other DAE studies examine how algebraic variables participate, both through modal participation factors and in singular systems more broadly \cite{TzounasEtAl2020,DassiosEtAl2020}. For large disturbances, critical clearing time sensitivities measure how parameters affect transient stability under a specific contingency \cite{Mishra2021}. The literature thus offers several useful importance measures. They answer different questions, though, and they can rank the same parameters differently.

\vspace{0.1cm}

\noindent\textbf{Parameter Identifiability.} Trajectory sensitivities support inverse problems and parameter estimation in nonlinear power system models \cite{Hiskens2004}, and structured identification recovers reduced-order models directly in differential-algebraic form \cite{NabaviChakrabortty2017}. \cite{LandgrafEtAl2022} study structural and practical identifiability for a nonlinear transient generator DAE, and \cite{EvangelouEtAl2024} compare Fisher information with data-driven approaches on a single machine infinite bus model. \cite{VirginilloEtAl2026} use a numerical Fisher information matrix to rank parameter estimators and informative disturbances for a small system. The closest precedent is our own recent study \cite{AlbustamiEtAl2026}, which develops DAE-aware Bayesian inference to jointly estimate generator inertia and damping along with transmission line resistance and reactance. That study does not connect identifiability to controllability, state observability, or stability-based parameter rankings.

\vspace{0.1cm}

\noindent\textbf{Research Gap.} Observability, controllability, parameter estimation, and stability sensitivity each have a large literature of their own. What is missing is a framework that keeps the full nonlinear generator and network DAE, evaluates all the properties in Tab.~\ref{tab:concepts} under many operating conditions, and turns the result into a ranking of parameter importance. No universal ranking should be expected, since every result depends on parameter scaling, sensors, actuators, excitation, time horizon, topology, operating point, and contingency. A useful outcome is instead a set of scenario-dependent rankings, together with an account of which parameters keep their rank across scenarios and of when the control-theoretic measures agree or conflict. Section~\ref{sec:case} reports both.

\vspace{0.1cm}

\noindent\textbf{Paper Contributions.}
The paper makes four contributions. \textit{(i)}~It casts parameter importance for nonlinear power system DAEs as one problem with four commensurately normalized score vectors, one each for controllability, identifiability, observability, and stability (Problem~\ref{prob:scores}). \textit{(ii)}~It builds DAE constraint-consistent variational Gramians, which extend the ODE variational Gramian of \cite{KazmaTaha2025} to the DAE manifold, and it shows that a continuous-time manifold reduction and a direct differentiation of the implicit time step give the same first-order output variation. \textit{(iii)}~It maps these Gramians to dimensionless metrics and then to scenario-aggregated parameter scores, which make explicit how each result depends on sensors, actuators, excitation, topology, horizon, and disturbance. \textit{(iv)}~It gives a scalable algorithm and tests it on multimachine benchmarks with different generator setups and model orders.

\vspace{0.1cm}

\noindent\textbf{Paper Organization.}
Section~\ref{sec:model} states the grid model and problem formulation. The variational dynamics follow in Section~\ref{sec:gramians}. Section~\ref{sec:scores} turns them into Gramians, property metrics, parameter scores, and the main algorithm, and Section~\ref{sec:case} tests the theory on three benchmark systems.

\section{Nonlinear DAE System Model}\label{sec:model}

Consider a transmission network with graph \(\mathcal{G}=(\mathcal{N},\mathcal{E})\), where \(\mathcal{N}=\{1,\ldots,N\}\) is the set of buses and \(\mathcal{E}\subseteq\mathcal{N}\times\mathcal{N}\) is the set of transmission lines. A bus can host synchronous generators, inverter-based resources, dynamic or static loads, and other controllable devices. The set \(\mathcal{N}_i=\{j\in\mathcal{N}\mid(i,j)\in\mathcal{E}\}\) collects the electrical neighbors of bus \(i\). Each branch \(\ell=(i,j)\in\mathcal{E}\) has a series conductance and susceptance, shunt elements, a transformer tap ratio, a phase shift, and a status. A topology is the active edge set \(\mathcal{E}(\boldsymbol{\sigma})\), where \(\boldsymbol{\sigma}\in\{0,1\}^{|\mathcal{E}_{\rm all}|}\) collects the status variables of the candidate branches \(\mathcal{E}_{\rm all}\). A line outage or a switching action changes \(\boldsymbol{\sigma}\) and, with it, the algebraic coupling. The complete nonlinear power system model then takes the DAE form
\begin{subequations}\label{eq:ndae}
\begin{align}
\dot{\mathbf{x}}(t)
  &=\mathbf{f}\bigl(\mathbf{x}(t),\mathbf{z}(t),\mathbf{u}(t),
      \mathbf{w}(t),\mathbf{p},\boldsymbol{\sigma}\bigr), \label{eq:ndae-d}\\
\mathbf{0}
  &=\mathbf{g}\bigl(\mathbf{x}(t),\mathbf{z}(t),\mathbf{u}(t),
      \mathbf{w}(t),\mathbf{p},\boldsymbol{\sigma}\bigr).\label{eq:ndae-a}
\end{align}
\end{subequations}
Tab.~\ref{tab:model-variables} lists the variables. The form \eqref{eq:ndae} is general enough to hold generator models of any order, along with the dynamics of exciters, governors, power system stabilizers, inverter controls, loads, and protection logic. Let \(\mathcal{N}_{\rm M}\subseteq\mathcal{N}\) be the buses equipped with PMUs. A PMU at bus \(i\in\mathcal{N}_{\rm M}\) measures the complex bus voltage and the complex currents on selected incident branches, and the aggregate measurement model is
\begin{equation}\label{eq:measurement}
\mathbf{y}(t)=
\mathbf{h}\bigl(\mathbf{x}(t),\mathbf{z}(t),\mathbf{u}(t),
                 \mathbf{p},\boldsymbol{\sigma}\bigr)
+\boldsymbol{\eta}(t),
\end{equation}
where \(\mathbf{y}\in\mathbb{R}^{n_y}\) collects the available PMU channels and \(\boldsymbol{\eta}\) is the measurement error. In rectangular coordinates a typical local output is \( \mathbf{y}_i= \operatorname{col}\!\left(
V_{Ri},V_{Ii}, \{I_{R,ij},I_{I,ij}\}_{j\in\mathcal{N}_i} \right).\) The nonlinear map \(\mathbf{h}\) encodes PMU placement, channel selection, coordinate transformations, branch currents computed from admittances, and any other measured device quantities. The parameter vector splits into blocks as
\begin{equation}\label{eq:p-vector}
\mathbf{p}
=\operatorname{col}\!\left(
\mathbf{p}_{\rm gen},\mathbf{p}_{\rm ctrl},\mathbf{p}_{\rm net},
\mathbf{p}_{\rm load},\mathbf{p}_{\rm res}
\right),
\end{equation}
which hold the generator, controller, network and power flow, load, and other resource parameters in that order. The list is meant to be inclusive, so the parameters of any device that appears in \(\mathbf{f}\), \(\mathbf{g}\), or \(\mathbf{h}\) become entries of \(\mathbf{p}\).

\begin{table}[t]
\centering
\caption{\normalfont Base vectors, dimensions, and representative entries, following \cite{SauerPaiChow2017}.}
\label{tab:model-variables}
\scriptsize
\tablestyle
\renewcommand{\arraystretch}{1.18}
\setlength{\tabcolsep}{3pt}
\begin{tabularx}{\columnwidth}{@{}>{\raggedright\arraybackslash}p{0.165\columnwidth}>{\raggedright\arraybackslash}p{0.2\columnwidth}>{\raggedright\arraybackslash}X@{}}
\toprule
\tabheadrow \tabh{Vector} & \tabh{Role} & \tabh{Representative entries} \\
\midrule
\(\mathbf{x}\in\mathbb{R}^{n_x}\) & Differential states & Generator \(\boldsymbol{\delta},\boldsymbol{\omega},\mathbf{E}'_q,\mathbf{E}'_d\), plus controller, inverter, storage, and dynamic load states \\
\(\mathbf{z}\in\mathbb{R}^{n_z}\) & Algebraic states & Bus voltages, stator currents, generated and consumed powers, and branch flows \\
\(\mathbf{u}\in\mathbb{R}^{n_u}\) & Control inputs & Mechanical power, field voltage, controller references, and inverter commands \\
\(\mathbf{w}\in\mathbb{R}^{n_w}\) & Uncertain inputs & Loads, renewable injections, faults, and other exogenous disturbances \\
\(\mathbf{y}\in\mathbb{R}^{n_y}\) & Outputs & PMU voltage and current channels and optional device measurements \\
\(\mathbf{p}\in\mathbb{R}^{n_p}\) & Parameters & \(M,D,R_s,x_d,x_q,x'_d,x'_q,T'_{d0},T'_{q0}\), controller gains and time constants, \(G_{ij},B_{ij}\), shunts, taps, and load and resource parameters \\
\(\boldsymbol{\sigma}\in\{0,1\}^{n_e}\) & Topology & Statuses of the \(n_e=|\mathcal{E}_{\rm all}|\) candidate branches \\
\(\mathbf{q}\in\mathbb{R}^{n_q}\) & Combined DAE vector & \(\mathbf{q}=\operatorname{col}(\mathbf{x},\mathbf{z})\), where \(n_q=n_x+n_z\) \\
\bottomrule
\end{tabularx}
\end{table}

The study specification \(\boldsymbol{\Omega}\) collects the admissible parameter ranges, consistent initial conditions, input and disturbance trajectories, PMU channels, actuators, time horizon, operating conditions, and topology scenarios. For every parameter \(p_k\) we seek four normalized scores \( s_{p_k}^{\rm c}, s_{p_k}^{\rm i}, s_{p_k}^{\rm o}, s_{p_k}^{\rm s}\in[0,1]\). The controllability score \(s_{p_k}^{\rm c}\) measures how strongly \(p_k\) changes the ability of the selected inputs to move the relevant states or modes. The identifiability score \(s_{p_k}^{\rm i}\) measures how uniquely and reliably \(p_k\) can be inferred from the prescribed input and measurement trajectories. Observability concerns the states, so \(s_{p_k}^{\rm o}\) measures how strongly \(p_k\) changes the information available for reconstructing the dynamic and algebraic states. Finally, the stability score \(s_{p_k}^{\rm s}\) measures the influence of \(p_k\) on the selected nonlinear stability margins. In every case zero means negligible relevance under \(\boldsymbol{\Omega}\), and one means the largest relevance after a stated normalization across the candidate parameters. This leads to the following problem.

\begin{problem}[Grid DAE parameter importance]\label{prob:scores}
Given the nonlinear DAE model \eqref{eq:ndae}, measurement model
\eqref{eq:measurement}, parameter vector \(\mathbf{p}\), and study specification
\(\boldsymbol{\Omega}\), compute the four score vectors \(\mathbf{s}^{\rm c},\
\mathbf{s}^{\rm i},\ \mathbf{s}^{\rm o},\ \mathbf{s}^{\rm s}\in[0,1]^{n_p}.\)
\end{problem}

The next section computes all four score vectors from \textit{variational} Gramians, following the steps in Fig.~\ref{fig:approach} which outlines the paper's methods and technical contributions.

\section{Variational Gramians for DAEs}\label{sec:gramians}
This section builds the variational sensitivities that feed the Gramians, the standard tools for assessing controllability and observability. Together they turn the nonlinear DAE trajectory into the four parameter scores of Fig.~\ref{fig:approach}.

\subsection{DAE Constraint-Consistent Variational Dynamics}
The goal is to see how the measured response changes when the initial state, the inputs, or the parameters change slightly. Following \cite{KazmaTaha2025}, we differentiate the nonlinear flow and its output map along a nominal trajectory. That trajectory always comes from solving the full nonlinear DAE \eqref{eq:ndae}. The variational equations describe only the small deviations around it, which are linear in the perturbation and have coefficients that change with time. The DAE adds one requirement that the ODE construction of \cite{KazmaTaha2025} lacks. The differential and algebraic variables must stay on the DAE constraint manifold \[ \mathcal{M}_{\mathbf{p},\boldsymbol{\sigma}}=
\left\{(\mathbf{x},\mathbf{z})\mid
\mathbf{g}(\mathbf{x},\mathbf{z},\mathbf{u},\mathbf{w},
\mathbf{p},\boldsymbol{\sigma})=\mathbf{0}\right\}. \] An algebraic perturbation that leaves this manifold violates the network equations and corresponds to \textit{no} nearby power system trajectory. We therefore perturb only the \(n_x\) differential states freely and let the algebraic variables follow from the manifold.

All Jacobians below are evaluated along the nominal nonlinear trajectory, and the dependence on \((\mathbf{u},\mathbf{w},\mathbf{p},\boldsymbol{\sigma})\) is dropped for compactness. The first variation is the \emph{directional derivative} of \eqref{eq:ndae} and \eqref{eq:measurement}, that is, the first-order change in the constrained trajectory and its output caused by an infinitesimal perturbation of the initial state, input, or parameter along a chosen direction. It reads
\begin{subequations}\label{eq:dae-variation}
\begin{align}
\delta\dot{\mathbf{x}}
 &=\mathbf{f}_{\mathbf{x}}\delta\mathbf{x}
   +\mathbf{f}_{\mathbf{z}}\delta\mathbf{z}
   +\mathbf{f}_{\mathbf{u}}\delta\mathbf{u}
   +\mathbf{f}_{\mathbf{p}}\delta\mathbf{p}, \label{eq:dae-var-d}\\
\textsc{\textbf{Var-DAE}}\;\;\; \hspace{0.5cm}\mathbf{0}
 &=\mathbf{g}_{\mathbf{x}}\delta\mathbf{x}
   +\mathbf{g}_{\mathbf{z}}\delta\mathbf{z}
   +\mathbf{g}_{\mathbf{u}}\delta\mathbf{u}
   +\mathbf{g}_{\mathbf{p}}\delta\mathbf{p}, \label{eq:dae-var-a}\\
\delta\mathbf{y}
 &=\mathbf{h}_{\mathbf{x}}\delta\mathbf{x}
   +\mathbf{h}_{\mathbf{z}}\delta\mathbf{z}
   +\mathbf{h}_{\mathbf{u}}\delta\mathbf{u}
   +\mathbf{h}_{\mathbf{p}}\delta\mathbf{p}. \label{eq:dae-var-y}
\end{align}
\end{subequations}
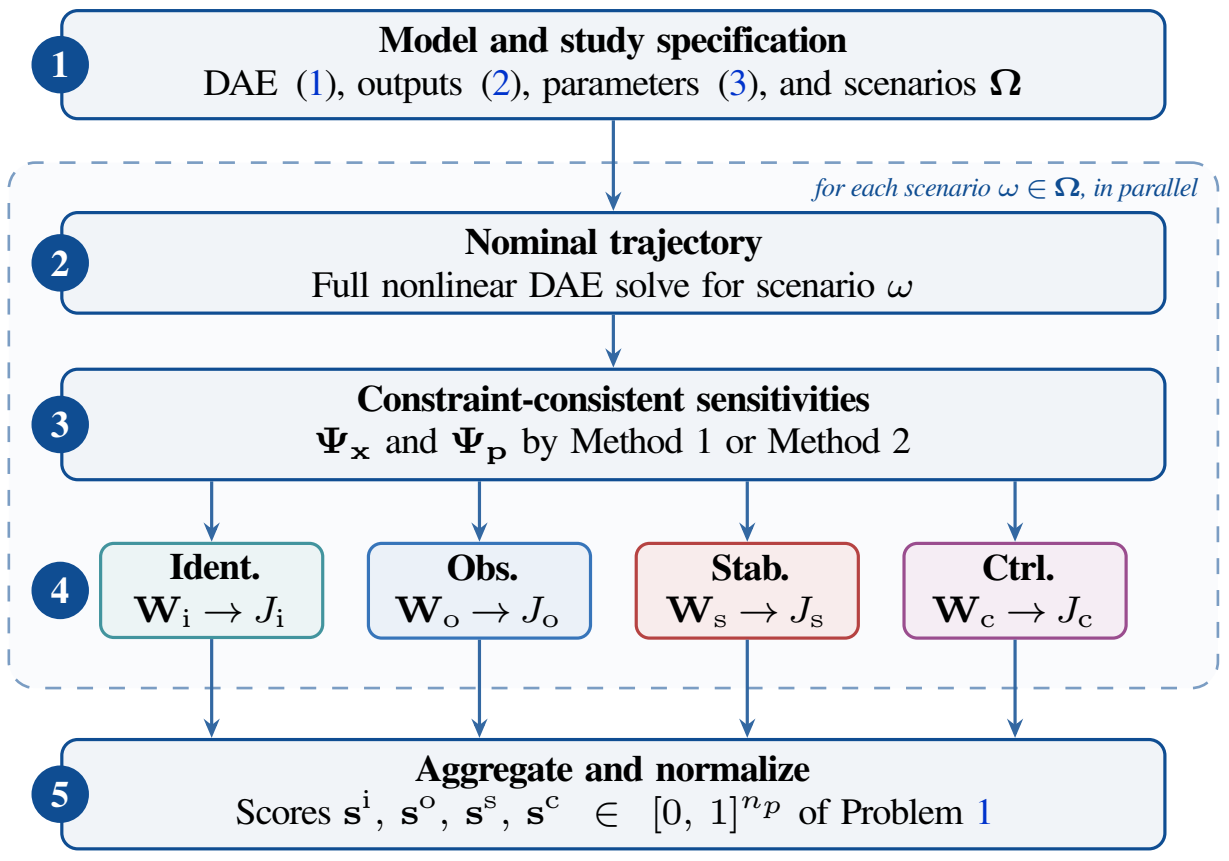
\begin{figure}[t]
\centering
\begin{tikzpicture}[node distance=3.5mm]
\node[stage] (s1) {\textbf{Model and study specification}\\ DAE \eqref{eq:ndae}, outputs \eqref{eq:measurement}, parameters \eqref{eq:p-vector}, and scenarios \(\boldsymbol{\Omega}\)};
\node[stage, below=6mm of s1] (s2) {\textbf{Nominal trajectory}\\ Full nonlinear DAE solve for scenario \(\omega\)};
\node[stage, below=of s2] (s3) {\textbf{Constraint-consistent sensitivities}\\ \(\boldsymbol{\Psi}_{\mathbf{x}}\) and \(\boldsymbol{\Psi}_{\mathbf{p}}\) by Method~1 or Method~2};
\node[gram, draw=f4pteal, fill=f4pteal!10, below=4mm of s3, xshift=-0.3\columnwidth] (gi) {\textbf{Ident.}\\ \(\mathbf{W}_{\rm i}\to J_{\rm i}\)};
\node[gram, draw=f4pblue2, fill=f4pblue2!10, below=4mm of s3, xshift=-0.1\columnwidth] (go) {\textbf{Obs.}\\ \(\mathbf{W}_{\rm o}\to J_{\rm o}\)};
\node[gram, draw=f4pred, fill=f4pred!10, below=4mm of s3, xshift=0.1\columnwidth] (gs) {\textbf{Stab.}\\ \(\mathbf{W}_{\rm s}\to J_{\rm s}\)};
\node[gram, draw=f4pviolet, fill=f4pviolet!10, below=4mm of s3, xshift=0.3\columnwidth] (gc) {\textbf{Ctrl.}\\ \(\mathbf{W}_{\rm c}\to J_{\rm c}\)};
\node[stage, below=6.5mm of gs, xshift=-0.1\columnwidth] (s5) {\textbf{Aggregate and normalize}\\ Scores \(\mathbf{s}^{\rm i},\mathbf{s}^{\rm o},\mathbf{s}^{\rm s},\mathbf{s}^{\rm c}\in[0,1]^{n_p}\) of Problem~\ref{prob:scores}};
\draw[flow] (s1) -- (s2);
\draw[flow] (s2) -- (s3);
\foreach \g in {gi,go,gs,gc} {
  \draw[flow] (\g.north |- s3.south) -- (\g.north);
  \draw[flow] (\g.south) -- (\g.south |- s5.north);
}
\begin{scope}[on background layer]
\node[loopbox, fit=(s2)(s3)(gi)(gc), inner xsep=3.4mm, inner ysep=3.2mm] (loop) {};
\end{scope}
\node[anchor=north east, font=\tiny\itshape, text=f4pblue, inner sep=1.3pt] at ([xshift=-2.5pt, yshift=-1.6pt]loop.north east) {for each scenario \(\omega\in\boldsymbol{\Omega}\), in parallel};
\node[badge] at (s1.west) {1};
\node[badge] at (s2.west) {2};
\node[badge] at (s3.west) {3};
\node[badge] at (s3.west |- gi.center) {4};
\node[badge] at (s5.west) {5};
\end{tikzpicture}
\caption{\normalfont Workflow for Problem~\ref{prob:scores}. Steps 2 to 4 run for every scenario \(\omega\in\boldsymbol{\Omega}\) in parallel, with one Gramian and one metric per property, and Step 5 aggregates and normalizes the scores.}
\label{fig:approach}
\end{figure}
We call \eqref{eq:dae-variation} the \emph{variational DAE} (\textsc{Var-DAE}). Each \(\delta\)-quantity is the perturbation of the matching variable, and the subscripted matrices are the Jacobians of \(\mathbf{f}\), \(\mathbf{g}\), and \(\mathbf{h}\) with respect to the states, algebraic variables, inputs, and parameters. The \textsc{Var-DAE} is a linearization taken \emph{along the full nonlinear trajectory}, so its coefficients vary in time and are recomputed at every point of the nonlinear solution. A linearization about a single equilibrium has constant coefficients instead. Equation~\eqref{eq:dae-variation} is the standard trajectory sensitivity formulation and is not new to this paper. Throughout, the DAE is regular, smooth, and of index one. These assumptions almost always hold for power system dynamics, so we state them only briefly. The vector-valued nonlinearities \(\mathbf{f}\), \(\mathbf{g}\), and \(\mathbf{h}\) are continuously differentiable, the initial condition is consistent, and \(\mathbf{g}_{\mathbf{z}}\) is nonsingular along the trajectory \cite{SauerPaiChow2017,GrossEtAl2016}. Two methods then quantify how each parameter affects the system properties, given in the next section.

\subsection{Constraint Manifold Reduction (Method 1)}

Under the regularity conditions above, the implicit function theorem applied to the algebraic block \eqref{eq:dae-var-a} of the variational DAE gives the constraint-consistent algebraic variation
\begin{equation}\label{eq:algebraic-elimination}
\delta\mathbf{z}=-\mathbf{g}_{\mathbf{z}}^{-1}
\left(\mathbf{g}_{\mathbf{x}}\delta\mathbf{x}
+\mathbf{g}_{\mathbf{u}}\delta\mathbf{u}
+\mathbf{g}_{\mathbf{p}}\delta\mathbf{p}\right).
\end{equation}
Substituting it into \eqref{eq:dae-variation} gives the reduced variational model
\begin{subequations}\label{eq:reduced-variation}
\begin{align}
\delta\dot{\mathbf{x}}
 &=\mathbf{A}\delta\mathbf{x}
   +\mathbf{B}\delta\mathbf{u}
   +\mathbf{E}\delta\mathbf{p},\\
\delta\mathbf{y}
 &=\mathbf{C}\delta\mathbf{x}
   +\mathbf{D}\delta\mathbf{u}
   +\mathbf{F}\delta\mathbf{p},
\end{align}
\end{subequations}
where
\begin{align*}
\mathbf{A}&=\mathbf{f}_{\mathbf{x}}-
\mathbf{f}_{\mathbf{z}}\mathbf{g}_{\mathbf{z}}^{-1}
\mathbf{g}_{\mathbf{x}},&
\mathbf{B}&=\mathbf{f}_{\mathbf{u}}-
\mathbf{f}_{\mathbf{z}}\mathbf{g}_{\mathbf{z}}^{-1}
\mathbf{g}_{\mathbf{u}},\\
\mathbf{E}&=\mathbf{f}_{\mathbf{p}}-
\mathbf{f}_{\mathbf{z}}\mathbf{g}_{\mathbf{z}}^{-1}
\mathbf{g}_{\mathbf{p}},&
\mathbf{C}&=\mathbf{h}_{\mathbf{x}}-
\mathbf{h}_{\mathbf{z}}\mathbf{g}_{\mathbf{z}}^{-1}
\mathbf{g}_{\mathbf{x}},\\
\mathbf{D}&=\mathbf{h}_{\mathbf{u}}-
\mathbf{h}_{\mathbf{z}}\mathbf{g}_{\mathbf{z}}^{-1}
\mathbf{g}_{\mathbf{u}},&
\mathbf{F}&=\mathbf{h}_{\mathbf{p}}-
\mathbf{h}_{\mathbf{z}}\mathbf{g}_{\mathbf{z}}^{-1}
\mathbf{g}_{\mathbf{p}}.
\end{align*}
The reduced matrices have dimensions \(\mathbf{A}\in\mathbb{R}^{n_x\times
n_x}\), \(\mathbf{B}\in\mathbb{R}^{n_x\times n_u}\),
\(\mathbf{E}\in\mathbb{R}^{n_x\times n_p}\),
\(\mathbf{C}\in\mathbb{R}^{n_y\times n_x}\),
\(\mathbf{D}\in\mathbb{R}^{n_y\times n_u}\), and
\(\mathbf{F}\in\mathbb{R}^{n_y\times n_p}\). In practice they come from sparse linear solves with \(\mathbf{g}_{\mathbf{z}}\), and its inverse is never formed. Let \(\boldsymbol{\Phi}(t,t_0)\in\mathbb{R}^{n_x\times n_x}\) be the state transition matrix and \(\mathbf{S}_{\mathbf{p}}(t)\in\mathbb{R}^{n_x\times n_p}\) the parameter sensitivity matrix. They satisfy
\begin{subequations}\label{eq:reduced-sensitivities}
\begin{align}
\dot{\boldsymbol{\Phi}}(t,t_0)
 &=\mathbf{A}(t)\boldsymbol{\Phi}(t,t_0),&
\boldsymbol{\Phi}(t_0,t_0)&=\mathbf{I}_{n_x},\\
\dot{\mathbf{S}}_{\mathbf{p}}(t)
 &=\mathbf{A}(t)\mathbf{S}_{\mathbf{p}}(t)+\mathbf{E}(t),&
\mathbf{S}_{\mathbf{p}}(t_0)&=\mathbf{S}_{\mathbf{p},0}.
\end{align}
\end{subequations}
The initial value \(\mathbf{S}_{\mathbf{p},0}\in\mathbb{R}^{n_x\times n_p}\) carries any dependence of the initial power flow and dynamic states on \(\mathbf{p}\). The output Jacobians, which are the sensitivities of the measured output \eqref{eq:measurement} to the initial state and to the parameters, are
\begin{equation}\label{eq:output-sensitivity}
\boldsymbol{\Psi}_{\mathbf{x}}(t)=\mathbf{C}(t)\boldsymbol{\Phi}(t,t_0),\;\;
\boldsymbol{\Psi}_{\mathbf{p}}(t)=\mathbf{C}(t)\mathbf{S}_{\mathbf{p}}(t)+\mathbf{F}(t),
\end{equation}
with \(\boldsymbol{\Psi}_{\mathbf{x}}(t)\in\mathbb{R}^{n_y\times n_x}\) and \(\boldsymbol{\Psi}_{\mathbf{p}}(t)\in\mathbb{R}^{n_y\times n_p}\). Method~1 thus eliminates the algebraic variables on the constraint manifold and integrates a reduced ordinary differential equation for the state-to-output and parameter-to-output sensitivities.

\subsection{DAE Time-Step Differentiation (Method 2)}

Method~2 computes the same first-order sensitivities as the variational DAE \eqref{eq:dae-variation}. It keeps \(\mathbf{x}\) and \(\mathbf{z}\) together and differentiates the discrete nonlinear equations that the time-domain integrator already solves at each step. Let \(\mathbf{q}_k=\operatorname{col}(\mathbf{x}_k,\mathbf{z}_k)\in\mathbb{R}^{n_q}\). With the input held over the step and the exogenous \(\mathbf{w}\) and topology \(\boldsymbol{\sigma}\) fixed at their nominal values, one integration step is the residual equation
\begin{equation}\label{eq:step-residual}
\mathbf{r}_k(\mathbf{q}_{k+1},\mathbf{q}_k,\mathbf{u}_k,\mathbf{p})=\mathbf{0},
\end{equation}
where \(\mathbf{r}_k\) maps \(\mathbb{R}^{2n_q+n_u+n_p}\) to \(\mathbb{R}^{n_q}\) and already embeds the algebraic constraints \(\mathbf{g}=\mathbf{0}\) at the new step. The residual can represent the trapezoidal rule that many power system simulators use, a backward differentiation formula, or an implicit Runge-Kutta method. Differentiating \eqref{eq:step-residual} along the same perturbation directions \((\delta\mathbf{q},\delta\mathbf{u},\delta\mathbf{p})\) gives
\begin{equation}\label{eq:discrete-tangent}
\delta\mathbf{q}_{k+1}
=\mathbf{A}^{\rm d}_k\delta\mathbf{q}_k
+\mathbf{B}^{\rm d}_k\delta\mathbf{u}_k
+\mathbf{E}^{\rm d}_k\delta\mathbf{p},
\end{equation}
with \(\mathbf{A}^{\rm d}_k=-\mathbf{R}_{k+1}^{-1}\mathbf{R}_k\),
\(\mathbf{B}^{\rm d}_k=-\mathbf{R}_{k+1}^{-1}\mathbf{R}_{\mathbf{u},k}\), and
\(\mathbf{E}^{\rm d}_k=-\mathbf{R}_{k+1}^{-1}\mathbf{R}_{\mathbf{p},k}\), where
\(\mathbf{R}_{k+1}=\partial\mathbf{r}_k/\partial\mathbf{q}_{k+1}\) and
\(\mathbf{R}_k=\partial\mathbf{r}_k/\partial\mathbf{q}_k\) belong to
\(\mathbb{R}^{n_q\times n_q}\),
\(\mathbf{R}_{\mathbf{u},k}\in\mathbb{R}^{n_q\times n_u}\), and
\(\mathbf{R}_{\mathbf{p},k}\in\mathbb{R}^{n_q\times n_p}\). Consequently,
\(\mathbf{A}^{\rm d}_k\in\mathbb{R}^{n_q\times n_q}\), \(\mathbf{B}^{\rm
d}_k\in\mathbb{R}^{n_q\times n_u}\), and \(\mathbf{E}^{\rm
d}_k\in\mathbb{R}^{n_q\times n_p}\). Here too the inverses stand for linear solves. The Newton solution of the nominal DAE step usually leaves a factorization of \(\mathbf{R}_{k+1}\) behind, so the tangent update is cheap. General-purpose DAE solvers use the same idea when they integrate forward sensitivities alongside the original DAE.

At \(t_0\) a perturbation of the differential states must be \emph{lifted} onto the constraint manifold. The lift completes it with the matching algebraic perturbation so that the pair satisfies \eqref{eq:dae-var-a}, and it reads
\begin{equation}\label{eq:tangent-lift}
\delta\mathbf{q}_0=\mathbf{T}_0\delta\mathbf{x}_0,\qquad
\mathbf{T}_0=
\begin{bmatrix}
\mathbf{I}\\
-\mathbf{g}_{\mathbf{z},0}^{-1}\mathbf{g}_{\mathbf{x},0}
\end{bmatrix}.
\end{equation}
Here \(\mathbf{T}_0\in\mathbb{R}^{n_q\times n_x}\) maps an independent differential-state perturbation into the tangent space of the DAE manifold. The product \(\boldsymbol{\Phi}^{\rm d}_{\mathbf{q},k} =\mathbf{A}^{\rm d}_{k-1}\cdots\mathbf{A}^{\rm d}_0\mathbf{T}_0\in\mathbb{R}^{n_q\times n_x}\) then maps a feasible perturbation of \(\mathbf{x}_0\) to \(\delta\mathbf{q}_k\), and the output variation is
\begin{equation}\label{eq:discrete-output}
\delta\mathbf{y}_k=
\mathbf{H}_k\delta\mathbf{q}_k+
\mathbf{H}_{\mathbf{u},k}\delta\mathbf{u}_k+
\mathbf{H}_{\mathbf{p},k}\delta\mathbf{p},
\end{equation}
with \(\mathbf{H}_k=[\,\mathbf{h}_{\mathbf{x},k}\ \
\mathbf{h}_{\mathbf{z},k}\,]\in\mathbb{R}^{n_y\times n_q}\),
\(\mathbf{H}_{\mathbf{u},k}\in\mathbb{R}^{n_y\times n_u}\), and
\(\mathbf{H}_{\mathbf{p},k}\in\mathbb{R}^{n_y\times n_p}\). Method~2 suits a power system simulation toolbox such as ANDES~\cite{CuiEtAl2021}, because it reuses the simulator's symbolic Jacobians, sparse Newton matrix, event handling, and numerical discretization. It also differentiates the numerical flow that the experiments actually use, whereas Method~1 differentiates the continuous DAE before discretization. Equations \eqref{eq:discrete-tangent} through \eqref{eq:discrete-output} form the discrete-time realization of the variational DAE \eqref{eq:dae-variation}. They act as a constraint-consistent propagator that carries an initial, input, or parameter perturbation through the simulated trajectory and out to the measurement. This propagator is the DAE counterpart of the variational dynamics that \cite{KazmaTaha2025} develop for ODE systems, and it respects the algebraic network constraints.

The two methods are equivalent, as the next proposition states. That lets us use the cheaper one, Method~2 with the simulator's own factorizations, without loss.

\begin{proposition}[Equivalence of the two DAE variations]
\label{prop:variation-equivalence}
The reduced variation of Method~1, \eqref{eq:reduced-variation}, and the
discrete variation of Method~2, \eqref{eq:discrete-tangent}, map the same DAE
constraint-consistent initial, input, and parameter perturbations to the same
first-order output variation in the limit as the integration step approaches
zero.
\end{proposition}

\noindent We omit the proof for brevity. Both constructions are the first derivative of the same constrained flow, and the case studies in Section~\ref{sec:case} confirm the equivalence numerically. The two methods therefore differ only in discretization error, which motivates a comparison of their accuracy, run time, conditioning, and time-domain behavior.

\section{Variational Metrics and Parameter Scores}\label{sec:scores}

This section turns the variational dynamics \eqref{eq:dae-variation} of Section~\ref{sec:gramians} into one Gramian per property, then into scalar metrics, and finally into normalized parameter scores.

\subsection{DAE Variational Gramians}

All four Gramians, for observability, identifiability, controllability, and stability, come from the same variational sensitivities on the DAE manifold.

\vspace{0.15cm}

\noindent\textbf{Observability.}
Let \(t_0,\ldots,t_K\) be the sample times of the output \eqref{eq:measurement}. The quadrature weights \(\alpha_k>0\), for example the trapezoidal weights \(\alpha_k=\tfrac12(t_{k+1}-t_{k-1})\), turn the sum below into a time integral over the horizon. The matrix \(\mathbf{R}_y\in\mathbb{R}^{n_y\times n_y}\), with \(\mathbf{R}_y\succ\mathbf{0}\), describes measurement noise or output scaling, \(\mathbf{D}_{\mathbf{x}}\in\mathbb{R}^{n_x\times n_x}\) is a nonsingular diagonal state-scaling matrix, and \(n_Y=(K+1)n_y\). Stacking the output-to-state sensitivity \(\boldsymbol{\Psi}_{\mathbf{x},k}=\boldsymbol{\Psi}_{\mathbf{x}}(t_k)\) from \eqref{eq:output-sensitivity} over all sample times gives the weighted sensitivity matrix \(\boldsymbol{\mathcal{V}}_{\rm o}\), whose block rows are the per-sample sensitivities, and the observability Gramian \(\mathbf{W}_{\rm o}\),
\begin{equation}\label{eq:dae-observability-gramian}
\boldsymbol{\mathcal{V}}_{\rm o}=
\begin{bmatrix}\sqrt{\alpha_0}\,\mathbf{R}_y^{-1/2}\boldsymbol{\Psi}_{\mathbf{x},0}\mathbf{D}_{\mathbf{x}}\\[-1pt]\vdots\\[-1pt]
\sqrt{\alpha_K}\,\mathbf{R}_y^{-1/2}\boldsymbol{\Psi}_{\mathbf{x},K}\mathbf{D}_{\mathbf{x}}\end{bmatrix},
\qquad
\mathbf{W}_{\rm o}=\boldsymbol{\mathcal{V}}_{\rm o}^{\mathsf{T}}\boldsymbol{\mathcal{V}}_{\rm o},
\end{equation}
with \(\boldsymbol{\mathcal{V}}_{\rm o}\in\mathbb{R}^{n_Y\times n_x}\) and \(\mathbf{W}_{\rm o}\in\mathbb{R}^{n_x\times n_x}\). Over the horizon, \(\mathbf{W}_{\rm o}\) accumulates how strongly each feasible initial-state direction shows up in the measurements, so a direction the sensors barely see adds little to it. It is the constraint-consistent DAE counterpart of the variational observability Gramian that \cite{KazmaTaha2025} build for ODE systems. The definition holds for either method of Section~\ref{sec:gramians}, and in Method~2 the sensitivity is \(\boldsymbol{\Psi}_{\mathbf{x},k}=\mathbf{H}_k\boldsymbol{\Phi}^{\rm d}_{\mathbf{q},k}\) from \eqref{eq:discrete-output}. The next proposition records when \(\mathbf{W}_{\rm o}\) certifies observability.

\begin{proposition}[DAE variational observability Gramian]\label{prop:dae-vargram}
The Gramian \(\mathbf{W}_{\rm o}\) in \eqref{eq:dae-observability-gramian} is
positive semidefinite, and positive definite if and only if the stacked output
Jacobian \(\boldsymbol{\mathcal{V}}_{\rm o}\) has full column rank in the
\(n_x\) independent tangent directions. Moreover, it is the infinitesimal limit
of an empirical observability Gramian formed by perturbing \(\mathbf{x}_0\) and
recomputing a consistent \(\mathbf{z}_0\) for every perturbed simulation.
\end{proposition}

\noindent We omit the proof for brevity, and the case studies in Section~\ref{sec:case} verify the result numerically.

\vspace{0.15cm}

\noindent\textbf{Identifiability.} The same stacking with the output-to-parameter sensitivity \(\boldsymbol{\Psi}_{\mathbf{p},k}=\boldsymbol{\Psi}_{\mathbf{p}}(t_k)\) from \eqref{eq:output-sensitivity} in place of the output-to-state sensitivity gives the identifiability matrix and Gramian
\begin{equation}\label{eq:identifiability-gramian}
\hspace{-0.3cm}\boldsymbol{\mathcal{V}}_{\rm i}=
\operatorname{stack}_{k=0}^{K}\!\big(\sqrt{\alpha_k}\,\mathbf{R}_y^{-1/2}
\boldsymbol{\Psi}_{\mathbf{p},k}\mathbf{D}_{\mathbf{p}}\big),
\;\;
\mathbf{W}_{\rm i}=\boldsymbol{\mathcal{V}}_{\rm i}^{\mathsf{T}}\boldsymbol{\mathcal{V}}_{\rm i},
\end{equation}
where \(\mathbf{D}_{\mathbf{p}}\in\mathbb{R}^{n_p\times n_p}\) is a nonsingular diagonal parameter-scaling matrix, \(\boldsymbol{\mathcal{V}}_{\rm i}\in\mathbb{R}^{n_Y\times n_p}\), and \(\mathbf{W}_{\rm i}\in\mathbb{R}^{n_p\times n_p}\). For Gaussian measurement noise, \(\mathbf{W}_{\rm i}\) is also the local Fisher information matrix. Its rank measures local parameter identifiability, and its conditioning reveals practical confounding between parameters. Uncertainty in the initial condition can enter as nuisance directions and be removed through a Schur complement.

\vspace{0.15cm}

\noindent\textbf{Controllability.} Let \(\boldsymbol{\Gamma}_{k,j}\in\mathbb{R}^{n_x\times n_u}\) be the discrete impulse response of the variation, which maps a unit actuator perturbation applied at time \(t_j\) to the resulting differential-state variation at \(t_k\). For the discrete model, \[
\boldsymbol{\Gamma}_{k,j}= \mathbf{L}_{\mathbf{x}} \mathbf{A}^{\rm
d}_{k-1}\cdots \mathbf{A}^{\rm d}_{j+1}\mathbf{B}^{\rm d}_{j}, \qquad j<k, \]
where \(\mathbf{L}_{\mathbf{x}}=[\,\mathbf{I}_{n_x}\ \mathbf{0}\,]\in\mathbb{R}^{n_x\times n_q}\) extracts the differential states. The controllability Gramian accumulates these responses over all injection times, so it measures how strongly the chosen actuators can move each feasible state direction over the horizon. A finite-horizon variational controllability Gramian is
\begin{equation}\label{eq:controllability-gramian}
\mathbf{W}_{\rm c}(t_k)=
\sum_{j=0}^{k-1}\beta_j
\widehat{\boldsymbol{\Gamma}}_{k,j}
\widehat{\boldsymbol{\Gamma}}_{k,j}^{\mathsf{T}},\;\;\;
\widehat{\boldsymbol{\Gamma}}_{k,j}=
\mathbf{D}_{\mathbf{x}}^{-1}
\boldsymbol{\Gamma}_{k,j}\mathbf{D}_{\mathbf{u}},
\end{equation}
where \(\mathbf{D}_{\mathbf{u}}\in\mathbb{R}^{n_u\times n_u}\) is a nonsingular diagonal input-scaling matrix, each \(\beta_j>0\) is a scalar weight, \(\widehat{\boldsymbol{\Gamma}}_{k,j}\in\mathbb{R}^{n_x\times n_u}\), and \(\mathbf{W}_{\rm c}(t_k)\in\mathbb{R}^{n_x\times n_x}\). This Gramian quantifies local reachability along the nonlinear constrained trajectory. Since \(\mathbf{A}^{\rm d}_k\) and \(\mathbf{B}^{\rm d}_k\) come from the step residual \eqref{eq:step-residual}, which embeds \(\mathbf{g}=\mathbf{0}\), the algebraic power flow constraints hold at every step, and the only reachable directions are those consistent with the network equations. Global nonlinear controllability is a separate question that \(\mathbf{W}_{\rm c}\) does not settle here, which could be an interesting (and frankly very challenging) question for future work. 

\vspace{0.15cm}

\noindent\textbf{Stability.} Stability is harder to pin down than the other three properties because it spans both small and large disturbances. We therefore use two complementary measures and say plainly what each one \textit{does} and \textit{does not certify}. The first is a local divergence measure built from the scaled finite-time Cauchy-Green matrix
\begin{equation}\label{eq:stability-deformation}
\mathbf{W}_{\rm s}(t_k)=
\widehat{\boldsymbol{\Phi}}_k^{\mathsf{T}}
\widehat{\boldsymbol{\Phi}}_k,\qquad
\widehat{\boldsymbol{\Phi}}_k=
\mathbf{D}_{\mathbf{x}}^{-1}
\mathbf{L}_{\mathbf{x}}\boldsymbol{\Phi}^{\rm d}_{\mathbf{q},k}
\mathbf{D}_{\mathbf{x}},
\end{equation}
where \(\widehat{\boldsymbol{\Phi}}_k\) and \(\mathbf{W}_{\rm s}(t_k)\) both belong to \(\mathbb{R}^{n_x\times n_x}\). The quantity \(\lambda_{\max}^{\rm FT}(t_k) =(2(t_k-t_0))^{-1}\log\lambda_{\max}(\mathbf{W}_{\rm s}(t_k))\) is the maximal finite-time Lyapunov exponent along the DAE manifold, and it measures how fast neighboring constrained trajectories separate over the horizon. Because \(\mathbf{W}_{\rm s}\) contains no measurement map, it is a \textit{deformation matrix} and not a Gramian in the strict sense. \cite{KazmaTahaLyap2025} study Lyapunov exponents of nonlinear DAE power systems in depth and optimize metrics over the full Lyapunov spectrum for stability assessment and renewable placement. 

Here a single finite-time exponent is one of two stability measures inside the parameter importance framework, and we refer the reader to that work for the full spectrum. The exponent is a local divergence rate. On its own it does not certify transient (large disturbance) stability, locate the stability boundary, or measure the margin of any particular contingency. A trajectory can have a modest finite-time exponent and still sit close to losing synchronism, and a large exponent does not always mean instability over the study horizon.

The second measure targets large disturbance margins directly. For a specified fault, the critical clearing time \(t_{\rm cc}(\mathbf{p})\) is the longest fault duration after which the post-fault DAE trajectory still returns to the acceptable operating region. Its parameter sensitivity
\begin{equation}\label{eq:cct-sensitivity}
\boldsymbol{\kappa}_{\ell}=
\frac{\partial t_{\rm cc}(\mathbf{p})}{\partial\log|p_\ell|}
\end{equation}
gives a contingency-specific stability importance that the finite-time exponent cannot provide. The critical clearing time sensitivity method for DAE models \cite{Mishra2021} computes it from trajectory sensitivities, with no linearization at an equilibrium. The first measure \eqref{eq:stability-deformation} thus captures local divergence along a generic trajectory, and the second \eqref{eq:cct-sensitivity} captures the margin of a named large disturbance.

\subsection{Property Metrics and Parameter Scores}

No single eigenvalue or trace captures every aspect of the four properties, so every reported ranking must state its metric. We use the regularized, dimensionless metrics
\begin{subequations}\label{eq:property-metrics}
\begin{align}
J_{\rm o}
 &=\log\det(\mathbf{W}_{\rm o}+\epsilon\mathbf{I}_{n_x}),\\
J_{\rm c}
 &=\log\det(\mathbf{W}_{\rm c}+\epsilon\mathbf{I}_{n_x}),\\
J_{\rm s}^{\rm div}
 &=\max_{1\leq k\leq K}\lambda_{\max}^{\rm FT}(t_k),\qquad
J_{\rm s}^{\rm cct}=t_{\rm cc}(\mathbf{p}),\\
J_{{\rm i},\ell}
 &=\left[(\mathbf{W}_{\rm i}
 +\epsilon\mathbf{I}_{n_p})^{-1}\right]_{\ell\ell}^{-1}.
\end{align}
\end{subequations}
For \(\epsilon>0\), all of \(J_{\rm o},J_{\rm c},J_{\rm s}^{\rm div},J_{\rm s}^{\rm cct}, J_{{\rm i},\ell}\in\mathbb{R}\). \(J_{\rm o}\) and \(J_{\rm c}\) reward information or control authority in every direction. \(J_{\rm s}^{\rm div}\) measures the worst case finite-time divergence and \(J_{\rm s}^{\rm cct}\) the large disturbance clearing margin, while \(J_{{\rm i},\ell}\) is the conditional information for parameter \(p_\ell\) after accounting for its correlation with the other parameters. The smallest eigenvalue, the trace, the condition number, and task-weighted variants are natural alternatives and robustness checks. A large trace, for instance, can hide a single nearly unobservable or uncontrollable direction.

Each \(\omega\in\boldsymbol{\Omega}\) is one study scenario, that is, one choice of operating point, disturbance, and topology from the specification \(\boldsymbol{\Omega}\) of Section~\ref{sec:model}. The operator \(\operatorname{Agg}_{\omega\in\boldsymbol{\Omega}}\) aggregates over those scenarios by a weighted mean, a worst case, or an upper quantile. Identifiability already has parameter-indexed columns, so its raw importance is the scenario-aggregated conditional information \[ r_{p_\ell}^{\rm i}=
\operatorname{Agg}_{\omega\in\boldsymbol{\Omega}} J_{{\rm i},\ell}(\omega). \]
For observability, controllability, and stability, a parameter is important when a dimensionless change in it substantially changes the corresponding property. Their raw scores are the aggregated elasticities
\begin{equation}\label{eq:raw-scores}
r_{p_\ell}^{q}=
\operatorname{Agg}_{\omega\in\boldsymbol{\Omega}}
\left|
\frac{\partial J_q(\omega)}
{\partial\log |p_\ell|}
\right|,\qquad q\in\{{\rm o},{\rm c}\}.
\end{equation}
The stability score uses the same construction summed over both stability metrics,
\begin{equation}\label{eq:raw-scores-stability}
r_{p_\ell}^{\rm s}=
\operatorname{Agg}_{\omega\in\boldsymbol{\Omega}}
\sum_{m\in\{{\rm div},{\rm cct}\}}
w_m\left|
\frac{\partial J_{\rm s}^{m}(\omega)}{\partial\log|p_\ell|}
\right|,
\end{equation}
where the elasticity of \(J_{\rm s}^{\rm cct}\) is exactly the critical clearing time sensitivity \(\boldsymbol{\kappa}_\ell\) of \eqref{eq:cct-sensitivity}. The weights \(w_m\ge0\) shift the emphasis between local divergence and the large disturbance margin, with \(w_{\rm cct}=0\) for disturbances that have no fault to clear. For a parameter that can be zero or change sign, the derivative is taken with respect to its declared characteristic scale in place of \(\log|p_\ell|\). Because \(J_{\rm o}\), \(J_{\rm c}\), and \(J_{\rm s}^{\rm div}\) already depend on first variations, their parameter derivatives are mixed sensitivities. Finite differences of the complete metric, second-order forward sensitivities, or an adjoint of the augmented nominal-plus-variational DAE can all compute them. The adjoint pays off when the number of scalar metrics is much smaller than \(n_p\) \cite{CaoEtAl2003}.

The final scores of Problem~\ref{prob:scores} are
\begin{equation}\label{eq:normalized-scores}
s_{p_\ell}^{q}=
\begin{cases}
r_{p_\ell}^{q}/\displaystyle\max_j \;\; r_{p_j}^{q},
 & \max_j \;\; r_{p_j}^{q}>0,\\
0, & \text{otherwise},
\end{cases}
\end{equation}
for \(q\in\{{\rm c},{\rm i},{\rm o},{\rm s}\}\). The normalization gives the largest raw score the value one and sets every score to zero in a zero-information case. Scores from different properties still live on different scales and are not numerically interchangeable. The sign of each elasticity in \eqref{eq:raw-scores} is kept separately to tell beneficial parameter changes from adverse ones.

The scores also have a limited scope. The Gramians and scores are local to a nonlinear trajectory and not to an equilibrium, and aggregation over \(\boldsymbol{\Omega}\) makes them aware of the operating region. Formal structural identifiability, global nonlinear controllability, and behavior near singular algebraic points need separate analysis, and a finite-horizon variational Gramian certifies none of them. With that in mind, seeing the methods are computationally cheap, they can arbitrarily be tested for thousands of scenarios. Algorithm~\ref{alg:workflow} summarizes the complete workflow of the paper.

\begin{algorithm}[t]
\caption{\normalfont NDAE Parameter Importance Scores}
\label{alg:workflow}
\SetAlgoVlined
\DontPrintSemicolon
\KwIn{DAE \eqref{eq:ndae}, PMU model \eqref{eq:measurement}, study specs
\(\boldsymbol{\Omega}\)}
\KwOut{Score vectors \(\mathbf{s}^{\rm i},\mathbf{s}^{\rm o},
\mathbf{s}^{\rm s},\mathbf{s}^{\rm c}\in[0,1]^{n_p}\)}

Set \(\mathbf{r}^{q}\leftarrow\mathbf{0}\),
\(q\in\{{\rm i},{\rm o},{\rm s},{\rm c}\}\)\;

\ForEach{\(\omega\in\boldsymbol{\Omega}\) \textbf{in parallel}}{
  Solve a consistent initial condition and the full nonlinear DAE
  \eqref{eq:ndae}, and split the trajectory at faults, switches, limits, and
  topology changes\;
  \ForEach{smooth segment}{
    Form the variational operators \eqref{eq:dae-variation} and factor
    \(\mathbf{g}_{\mathbf{z}}\) (Method~1) or \(\mathbf{R}_{k+1}\) (Method~2)
    once per step\;
    Propagate the sensitivities by Method~1
    \eqref{eq:reduced-variation}--\eqref{eq:output-sensitivity} or Method~2
    \eqref{eq:discrete-tangent}--\eqref{eq:tangent-lift}\;
    Accumulate \(\mathbf{W}_{\rm o}\) \eqref{eq:dae-observability-gramian},
    \(\mathbf{W}_{\rm i}\) \eqref{eq:identifiability-gramian},
    \(\mathbf{W}_{\rm c}\) \eqref{eq:controllability-gramian}, and
    \(\widehat{\boldsymbol{\Phi}}_k\) \eqref{eq:stability-deformation}\;
    Apply event-time and reset sensitivity maps\;
  }

  Evaluate \(J_{\rm i},J_{\rm o},J_{\rm s},J_{\rm c}\)
  \eqref{eq:property-metrics}\;
  Take \(r^{\rm i}_{p_\ell}\) from \(\mathbf{W}_{\rm i}\) and obtain the O/S/C
  elasticities \eqref{eq:raw-scores} by centered full-DAE differences\;
  Accumulate \(\mathbf{r}^{\rm i},\mathbf{r}^{\rm o},
  \mathbf{r}^{\rm s},\mathbf{r}^{\rm c}\)\;
}

Normalize each \(\mathbf{r}^{q}\) by \eqref{eq:normalized-scores}\;
\Return \(\mathbf{s}^{\rm i},\mathbf{s}^{\rm o},
\mathbf{s}^{\rm s},\mathbf{s}^{\rm c}\)\;
\end{algorithm}

\section{Case Studies}\label{sec:case}

\subsection{Simulation Setup and Research Questions}

All studies run on an M5 Max MacBook Pro (2026) with 128 GB of RAM. We implement Algorithm~\ref{alg:workflow} in ANDES, an open-source Python library for power system modeling and time-domain simulation \cite{CuiEtAl2021}.\footnote{ This toolbox, by the way, deserves more credit and should be used more often in power system studies. It can run ultra fast DAE simulations for arbitrary models of power systems and is very modular, and well-built, before the age of LLMs made programming cheap. } The test systems, in order of size, are the WSCC 9-bus system (\emph{case9}), the IEEE 39-bus New England system (\emph{case39}), and the WECC 179-bus system. ANDES supplies the nonlinear DAE \eqref{eq:ndae}, its trajectory, and the sparse DAE Jacobians that the variational methods use, and the machine, controller, and network data come from its distribution. The 9-bus model has three sixth-order machines and no controllers. The 39-bus and 179-bus models add turbine governors, exciters, and stabilizers, and Tab.~\ref{tab:timing} lists the state and algebraic dimensions \(n_x\) and \(n_z\) of all three. In every system the measurements are generator rotor speeds and terminal voltage magnitudes. The candidate parameters include the machine quantities \(M,D,R_s,x_d,x_q,x'_d,x'_q,T'_{d0},T'_{q0}\) and also the network power flow parameters that enter the algebraic equations, which are the transmission line series resistance \(r_{ij}\), series reactance \(x_{ij}\), and line charging \(b_{ij}\). Since the framework acts on the full DAE, it scores these algebraic (network) parameters on the same footing as the differential (machine) ones. Four research questions (RQs) guide the case studies.
\begin{itemize}\setlength\itemsep{0pt}\setlength\parskip{0pt}\setlength\topsep{1pt}
\item \textbf{RQ1. }Is the variational approach correct, i.e., do Methods~1 and~2 produce identical first-order sensitivities
(Prop.~\ref{prop:variation-equivalence}), and does the variational observability
Gramian match the empirical one (Prop.~\ref{prop:dae-vargram})?
\item\textbf{RQ2. } Do the four control-theoretic properties induce the same
parameter ranking, or do they conflict?
\item \textbf{RQ3. } How do the rankings depend on the disturbance and operating
point?
\item \textbf{RQ4. }What is the offline computational cost of a complete study,
and how does it grow with system size?
\end{itemize}

\subsection{Theory Validation (RQ1)}

\begin{figure*}[t]
\centering
\includegraphics[width=0.92\textwidth]{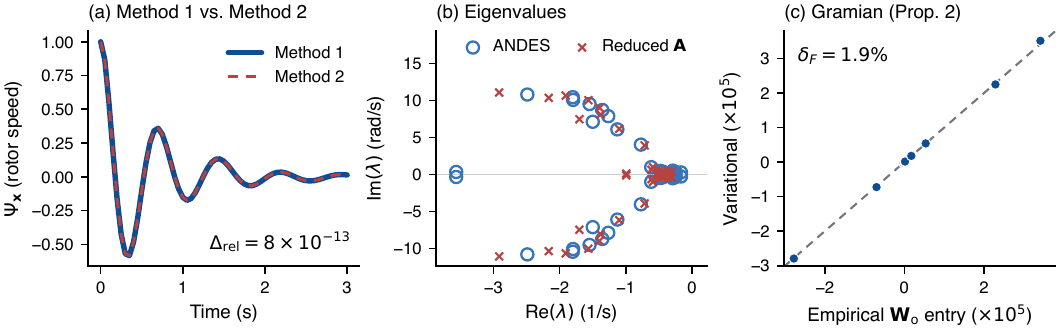}
\caption{\normalfont Validation of the theory (RQ1). (a)~On the IEEE 39-bus system, the output sensitivities
\(\boldsymbol{\Psi}_{\mathbf{x}}\) from Method~1 and Method~2 are indistinguishable
(\(\Delta_{\rm rel}\!\approx\!10^{-12}\)), as Proposition~\ref{prop:variation-equivalence} predicts. (b)~The eigenvalues of the reduced
state matrix \(\mathbf{A}=\mathbf{T}_f^{-1}(\mathbf{f}_{\mathbf{x}}-
\mathbf{f}_{\mathbf{z}}\mathbf{g}_{\mathbf{z}}^{-1}\mathbf{g}_{\mathbf{x}})\)
coincide with those of the independent ANDES eigenvalue routine. (c)~The variational
\emph{observability} Gramian \(\mathbf{W}_{\rm o}\) matches the empirical one
\(\mathbf{W}_{\rm o}^{\rm emp}\) built from full nonlinear resimulations
(Proposition~\ref{prop:dae-vargram}). Their relative Frobenius difference
\(\delta_F=\|\mathbf{W}_{\rm o}-\mathbf{W}_{\rm o}^{\rm emp}\|_F/
\|\mathbf{W}_{\rm o}\|_F\approx1.9\%\) comes from the finite-difference step.}
\label{fig:validation}
\end{figure*}

Fig.~\ref{fig:validation} answers RQ1 with three tests. First, Method~1 (manifold reduction, \eqref{eq:reduced-variation}) and Method~2 (direct differentiation of the implicit step, \eqref{eq:discrete-tangent}) give output sensitivities that agree to about \(10^{-12}\) over the whole horizon. Both discretize the same trajectory with the same integrator, so the equivalence of Proposition~\ref{prop:variation-equivalence} holds to round-off and not only in the small-step limit. Second, the eigenvalues of the reduced state matrix coincide with those from the separate ANDES eigenvalue analysis. They reproduce the electromechanical modes from \(0.2\) to \(2.3\)~Hz, which checks the constraint-consistent reduction against an external ground truth. The agreement holds on every test system, including the \(708\)-state WECC model, so the reduction stays accurate at scale. Third, the variational observability Gramian agrees with the empirical Gramian assembled from constraint-consistent perturbations of the initial state and full nonlinear resimulations. Their \(1.9\%\) relative Frobenius difference comes from the finite-difference step, and this comparison is the direct check of Proposition~\ref{prop:dae-vargram}. One numerical detail needs care. On the IEEE 39-bus system, \(50\) controller states of the exciters and stabilizers are written in algebraic form and make the bare algebraic block rank deficient. A small singular-perturbation mass keeps \(\mathbf{g}_{\mathbf{z}}\) nonsingular and reproduces the ANDES modes to within \(0.3\%\), and Method~2 stays well posed throughout.

The variational construction is also much cheaper than the empirical (finite-difference) Gramian it matches. The empirical route needs a full nonlinear resimulation for every perturbed direction, while the variational route reuses one trajectory and its factorizations. Tab.~\ref{tab:gramian-timing} compares the time per Gramian of the two routes on the IEEE 39-bus system.

\begin{table}[t]
\centering
\caption{\normalfont Time per Gramian (s) on the IEEE 39-bus system for the empirical route (finite-difference resimulation) and the variational route, both on the same single scenario and horizon.}
\label{tab:gramian-timing}
\footnotesize
\tablestyle
\begin{tabularx}{\columnwidth}{@{}l*{3}{>{\raggedleft\arraybackslash}X}@{}}
\toprule
\tabheadrow \tabh{Gramian} & \tabh{Empirical (s)} & \tabh{Variational (s)} & \tabh{Speedup} \\
\midrule
Observability & 81.8 & 0.24 & \textbf{345\(\times\)} \\
Controllability & 1.3 & 0.29 & \textbf{4\(\times\)} \\
Stability & 81.8 & 0.24 & \textbf{348\(\times\)} \\
\bottomrule
\end{tabularx}

\end{table}

\subsection{Conflicting Rankings (RQ2)}

\begin{figure}[t]
\centering
\includegraphics[width=0.95\columnwidth]{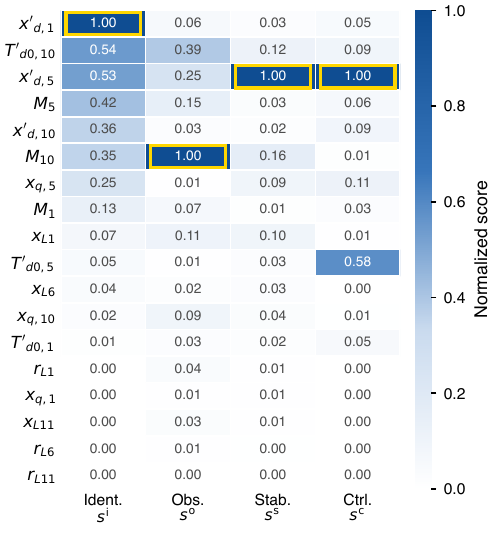}
\caption{\normalfont Normalized parameter scores on the IEEE 39-bus system, with rows ordered by identifiability and a candidate set that mixes machine and network line parameters. Gold boxes mark the top parameter of each property. A transient reactance \(x'_{d,1}\) leads identifiability, an inertia \(M_{10}\) leads observability, and a second transient reactance \(x'_{d,5}\) leads both stability and controllability. Damping \(D\) and stator resistance \(R_s\) are negligible throughout.}
\label{fig:scores}
\end{figure}

\begin{figure}[t]
\centering
\includegraphics[width=0.95\columnwidth]{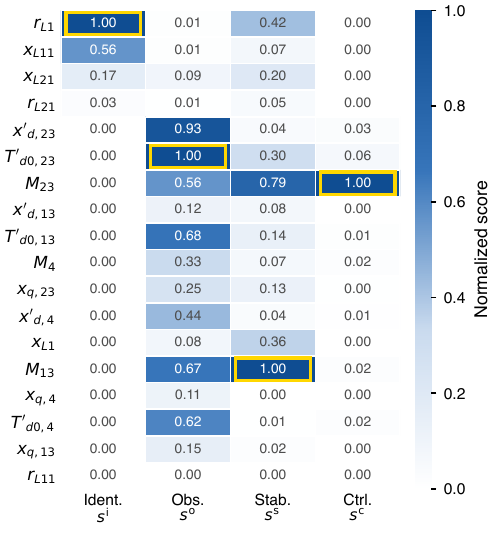}
\caption{\normalfont All four scores on the \(708\)-state WECC 179-bus system for one representative contingency, with controllability from the variational Gramian. The conflict persists at scale. A transmission line resistance \(r_{L1}\) tops identifiability and a field time constant \(T'_{d0,23}\) tops observability, while two different inertias, \(M_{13}\) and \(M_{23}\), top stability and controllability. Because a \emph{network} parameter leads one property, network power flow parameters belong in the candidate set.}
\label{fig:scores-wecc}
\end{figure}

\begin{figure*}[t]
\centering
\includegraphics[width=0.92\textwidth]{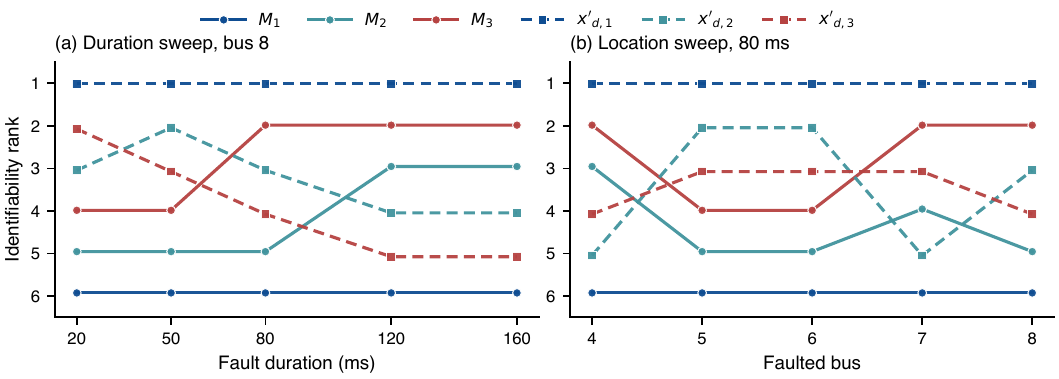}
\caption{\normalfont Scenario dependence of identifiability on the WSCC 9-bus system for the inertias \(M\) (solid) and transient reactances \(x'_d\) (dashed) of the three machines, with color marking the machine. The sweeps run over fault duration~(a) and fault location~(b), and rank 1 is the most identifiable. Both extremes stay fixed in either sweep. The slack-machine transient reactance \(x'_{d,1}\) is always first and the slack-machine inertia \(M_1\) is always last. The four parameters in between reorder by a similar amount in both panels (mean pairwise Kendall \(\tau=0.63\) in each), so the ranking separates scenario-independent parameters from scenario-dependent ones, and fault duration and fault location reshuffle it about equally.}
\label{fig:faultsweep}
\end{figure*}

Fig.~\ref{fig:scores} shows the normalized scores \eqref{eq:normalized-scores} for the IEEE 39-bus system, with machine and network line parameters in the same candidate set. The properties do not agree on a leader. A machine transient reactance \(x'_d\) dominates identifiability and an inertia \(M\) dominates observability, while a second transient reactance dominates both stability and controllability, for three distinct winners across four properties. The four control-theoretic measures thus disagree on which parameters carry the most weight, as the introduction anticipates. No universal ranking exists, and calibrating the most identifiable parameter does little for the parameters that govern dynamic security. Damping \(D\) and stator resistance \(R_s\) score near zero under every property, which fits the absence of a damping torque path and the negligible resistive losses in the chosen measurements. The disagreement persists on the \(708\)-state WECC system (Fig.~\ref{fig:scores-wecc}). There a \emph{network} parameter, a transmission line resistance, is the most identifiable, and a field time constant and two inertias lead observability, stability, and controllability. System size does not cause the conflict, and network power flow parameters can carry the most dynamic leverage.

\subsection{Scenario Dependence (RQ3)}
Fig.~\ref{fig:faultsweep} varies the disturbance on the 9-bus system, sweeping the fault duration in~(a) and the faulted bus in~(b). The ranking splits the candidate parameters into two groups. Two of them never move. The slack-machine transient reactance \(x'_{d,1}\) is the most identifiable parameter in every scenario of both sweeps, and the slack-machine inertia \(M_1\) is the least identifiable in every scenario. The slack machine thus holds both extremes, because the measured speed and voltage response is most sensitive to how this machine couples to the network and least sensitive to how fast it accelerates. The other four parameters keep trading places, and they do so to a similar degree in both sweeps. The mean pairwise Kendall rank correlation across the five scenarios is \(0.63\) for the duration sweep and also \(0.63\) for the location sweep. Fault duration and fault location therefore reorder the middle of the ranking about equally, and neither one is the gentler perturbation. This is the scenario-dependent behavior that the introduction anticipates, with one refinement. Rank persistence belongs to \emph{particular parameters} and not to \emph{particular kinds of disturbance}. Some parameters stay informative in every experiment, and others are informative only under specific disturbances, which argues for designing identification experiments around the parameters of interest.

\subsection{Computational Cost (RQ4)}
The framework is an \emph{offline} assessment that runs ahead of operation for planning, calibration, and model validation. The relevant cost is therefore the end-to-end time of a complete study, and online latency plays no role. Tab.~\ref{tab:timing} reports that time for all four score vectors over \(|\boldsymbol{\Omega}|=5\) scenarios. The variational controllability Gramian reuses the trajectory of the other scores, so no score needs an extra simulation. The 9-bus and 39-bus studies finish in about \(75\)~s each, and the \(708\)-state WECC study takes roughly eight minutes. The variational route is what keeps a study of this size tractable. By finite-difference resimulation a single observability Gramian takes more than a minute, while the variational route needs a fraction of a second (Tab.~\ref{tab:gramian-timing}).

\begin{table}[t]
\centering
\caption{\normalfont Computational time of Algorithm~\ref{alg:workflow} for all four score
vectors (offline study, variational controllability). \(n_d\) is the number of
differential states with a nonzero time constant.}
\label{tab:timing}
\footnotesize
\tablestyle
\begin{tabularx}{\columnwidth}{@{}l*{6}{>{\raggedleft\arraybackslash}X}@{}}
\toprule
\tabheadrow \tabh{System} & \tabh{\(n_x\)} & \tabh{\(n_d\)} & \tabh{\(n_z\)} & \tabh{\(n_p\)} & \tabh{\(|\boldsymbol{\Omega}|\)} & \tabh{Time (s)} \\
\midrule
WSCC 9 & 18 & 18 & 69 & 18 & 5 & \textbf{75.4} \\
IEEE 39 & 220 & 170 & 479 & 18 & 5 & \textbf{76.6} \\
WECC 179 & 708 & 573 & 1696 & 18 & 5 & \textbf{460.3} \\
\bottomrule
\end{tabularx}

\end{table}

\subsection{Discussion and Broader Impact}
The studies answer the four questions and support the central premise of the paper. The variational machinery passes every check in RQ1. The two methods agree to round-off, the variational and empirical observability Gramians match to about \(1.9\%\), and the reduced state matrix reproduces an independent eigensolver on all systems up to \(708\) states. The four properties then rank parameters differently on both the IEEE 39-bus and WECC systems (RQ2). Parameter importance is thus defined only relative to a stated purpose, a sensor and actuator set, and a disturbance, and a calibration campaign tuned for one property may leave another poorly constrained. Under RQ3 the rankings reorder with both fault duration and fault location, yet a few parameters hold their position in every scenario. This separates parameters that are informative everywhere from those that are informative only under specific disturbances, which can guide experiment and measurement design. Finally, the algorithm delivers all four scores over five scenarios on a \(708\)-state model in about eight minutes, orders of magnitude faster than the empirical route (RQ4).

The scores tell an operator which parameters to calibrate for a trustworthy model, where to place measurements, and which actuators to rely on for control and security, so calibration effort no longer has to treat all parameters as equally important. Future work will add structural identifiability certificates, analytic sensitivity propagation that avoids finite-difference parameter perturbations, inverter-based resource models, and larger systems.

\section*{Acknowledgments}
The author acknowledges support from the National Science Foundation (NSF), United States, under grants 2152450 and 2550934. The author also acknowledges the use of Claude to help develop and refine the theoretical results (the proofs were omitted but confirm the theory is correct) and with the code generation and implementation. The case studies and all data are available to any interested reader.

\bibliographystyle{IEEEtran}
\bibliography{refs}

\end{document}